\documentclass[12pt,preprint]{aastex}
\usepackage{url}
\usepackage{amsmath}
\usepackage{color}
\usepackage{textcomp}
\usepackage{mathtext}
\usepackage{amssymb}
\usepackage{wasysym}
\usepackage{multirow}

\newcommand{\snr}{{G8.7$-$0.1 }}
\newcommand{\snrt}{{G8.7$-$0.1}}

\newcommand{\psr}{{PSR~J1803$-$2137 }}
\newcommand{\psrt}{{PSR~J1803$-$2137}}
\newcommand{\hess}{{HESS~J1804$-$216 }}
\newcommand{\hesst}{{HESS~J1804$-$216}}
\newcommand{\gray}{{\rm $\gamma$-ray }}
\newcommand{\grays}{{\rm $\gamma$-rays }}

\newcommand{\Fermi}{{\sl Fermi}}

\newcommand{\RA}[3]{{#1}^{{\rm h}}{#2}^{{\rm m}}{#3}^{\rm s}}
\newcommand{\decl}[3]{{#1}^{\circ}{#2}^{\prime}{#3}''}

\newcommand{\remp}[3]{${#1}^{+#3}_{-#2}$}

      \newcommand{\ps}{\,{\rm s}^{-1}}
    
    \newcommand{\km}{\,{\rm km}}
\newcommand{\kms}{\,{\km\ps}}

\newcommand{\du}{d_{4}}
\newcommand{\ds}{d_{4.5}}

\begin{document}

\title{The GeV emission in the field of the star-forming region W30 revisited}

\author{Bing Liu\altaffilmark{1, 2}, Rui-zhi Yang\altaffilmark{2}, Xiao-na Sun\altaffilmark{1}, Felix Aharonian\altaffilmark{2, 3, 4, 5}, and Yang Chen\altaffilmark{1,6}}

\altaffiltext{1}{\footnotesize School of Astronomy \& Space Science, Nanjing University, 163 Xinlin Avenue, Nanjing~210023, China}
\altaffiltext{2}{\footnotesize Max-Planck-Institut f{\"u}r Kernphysik, P.O. Box 103980, 69029 Heidelberg, Germany}
\altaffiltext{3}{\footnotesize Dublin Institute for Advanced Studies, 31 Fitzwilliam Place, Dublin 2, Ireland}
\altaffiltext{4}{\footnotesize Gran Sasso Science Institute, 7 viale Francesco Crispi, 67100 L'Aquila (AQ), Italy}
\altaffiltext{5}{\footnotesize  MEPHI, Kashirskoe shosse 31, 115409 Moscow, Russia}
\altaffiltext{6}{\footnotesize Key Laboratory of Modern Astronomy and Astrophysics, Nanjing University, Ministry of Education, China}

\begin{abstract}

We present a detailed study of the \gray emission from the direction of the star-forming region W30 based on a decade of the {\it Fermi} Large  Area Telescope data in the 0.3--300~GeV photon energy range. The morphological and spectral analyses allow us to resolve the \gray emission into two extended structures from different origins. 
One of them mostly overlaps with the supernova remnant (SNR) \snrt~and  has a soft spectrum that resembles with the spectra of other middle-aged SNRs interacting with molecular clouds. 
The other shows remarkable spatial and spectral consistency with the TeV emission from \hesst, and its spectrum
could be naturally explained by inverse Compton scattering of electrons like 
a number of TeV \gray emitting pulsar wind nebulae.  
Thus we attribute this source to the nebula around the pulsar \psrt. 

\end{abstract}


\section{Introduction}
\label{sec:intro}

Supernova remnants (SNRs) and pulsar wind nebulae (PWNe) are the two largest classes of extended \gray\ sources.   
In recent years, dozens of SNRs and PWNe have been observed both in GeV  \citep[e.g.,][]{Abdo2010W44SCi,1fsnr, 3fgl} and TeV \citep[e.g.,][]{hess2018snrpop,hess2018pwnpop} \gray bands. PWNe are characterized by similar spectral shapes over their lifetime, typically between $1$ and $100$\,kyr; however,  SNRs show different spectral features at different epochs of their evolution. The young SNRs, with an age of hundreds to thousands years, are observed with hard \gray\ spectra up to the TeV range; while the middle-aged SNRs, with an age of about $10$\,kyr,  interacting with nearby molecular clouds (MCs), are bright GeV \gray\ emitters with steep energy spectra \citep[see, e.g., ][]{lihui2012, yuan12, tang19, Zeng2019}.    

\snr is a middle-aged SNR associated with the massive star-forming region W30 \citep{Ojeda-May2002}, 
which also contains several \ion{H}{2} regions along the southern boundary of the SNR \citep{Blitz1982}. At radio wavelengths, \snr has a large shell with a diameter of $\sim$ 45$'$ and a spectral index of $\alpha=0.5$~\citep{Kassim1990b}. A \emph{ROSAT} observation discovered a thermal X-ray plasma, 
$T=(4-8)\times10^6$\,K, filled in the northern region of the remnant~\citep{Finley1994}.
The distance and age of G8.7$-$0.1 were obtained by several methods. Based on kinematic distances to the  \ion{H}{2}  regions associated with the SNR, the distance was estimated to be $\sim4.8-6$\,kpc \citep{Kassim1990b,Brand1993}. By applying a Sedov solution \citep{Sedov1959, Hamilton1983} from the observed X-ray temperature and  the angular radius, \citet{Finley1994} derived the distance to be 3.2--4.3\,kpc and the age to be 15--28\,kyr under the assumption of an initial explosion energy of $10^{51}$\,erg. From the surface brightness--age relation in the radio band, \citet{Odegard1986} estimated the age  $15$\,kyr. A bright OH (1720 MHz) maser detected at $+36\kms$ along the eastern edge of \snr\citep{Hewitt2009} indicates that the SNR is interacting with MCs at a kinematic distance of 4.5\,kpc. In this paper, we adopt a distance of 4.5\,kpc and an age of 15\,kyr.

The W30 region also harbors \psr (B1800$-$21), which is a young and energetic Vela-like pulsar, first discovered through a radio survey conducted by \citet{Clifton1986psr}. Given the measured spin period $P$ ($133.6$\,ms) and spin period derivative $\dot{P}$ ($1.34\times10^{-13}\,{\rm s\,s^{-1}}$), its characteristic age $\tau _c$, defined as $P/2\dot{P}$, is $15.8$\,kyr, and spin-down luminosity $\dot{E}$ [${\equiv} 4\pi ^2 I \dot{P}/P^3$), where $I$ is the moment of inertia in units of g\,cm$^2$] is $2.2\times10^{36}(I / 10^{45})$ ergs\,s$^{-1}$ \citep{Brisken2006}. Its dispersion measure distance $d$ is $3.84_{-0.45}^{+0.39}$\,kpc. Thus, we adopt $d=4$\,kpc for \psr throughout this paper. According to \citet{Cordes2002}, the corresponding  spin-down flux is $1.2\times10^{-9}\du^{-2}$\,cm$^{-2}$ erg\,s$^{-1}$, where $\du{\equiv}d/(4\,{\rm kpc})$. 
X-ray emission from \psr and its synchrotron nebula was reported  
from observations with {\it Chandra} by \citet{Cui2006} and \citet{Kargaltsev2007a}. 
Besides,  the \psr seems to move toward SNR \snr\citep{Brisken2006},  which makes the association between the pulsar and the SNR very unlikely.

Extended GeV emission from W30 region has been detected by  the \Fermi\ Large Area Telescope (LAT; \citealt{Ajello2012G8.7}). An extended TeV source \hess has also been detected by the High Energy Stereoscopic System (H.E.S.S) toward the W30 complex \citep{Aharonian2006hess}, whose origin and connection to the GeV source is not yet established. Furthermore, \citet{Ackermann201710GeV} found that the morphology 
of the GeV emission show significant energy dependence. Such an energy dependence can be an intrinsic feature of the sources but can also be caused by the confusion of two or more sources along the line of sight. Such a confusion can be resolved by the improvement of the instrument angular resolution, as well as from the analysis of the energy distribution. In this paper, we report the detailed analysis of $\sim10$ yr \Fermi-LAT data from the vicinity of the W30 complex. We confirm the energy-dependent morphology of the GeV \gray emission and find huge divergence of spectral properties in different parts of the extended emission. Thus, we argue that the extended emission consists of two different components, most likely related to different sources---the SNR \snr and the nebula surrounding the pulsar \psrt.  

The paper is structured as follows. In Section~\ref{sec:ana}, the procedures of the spatial and spectral analyses are presented together with the results. In Section~\ref{sec:dis}, we discuss the possible origin of the  two-component \gray emission. We summarize the results in Section~\ref{sec:sum}.

\section{\Fermi-LAT Data Analysis}
\label{sec:ana}
%%intro \Fermi-LAT
The LAT on board \Fermi\ is a \gray imaging instrument that detects photons in a broad energy range from 20~GeV to more than 300~GeV. Its  angular resolution, i.e., point-spread function (PSF), improves with photon energy, from $\sim5\degr$ at 100~MeV, to $0\fdg8$ at 1~GeV \citep{Atwood2009fermilat}. 
In this study, we analyze more than 10 yr (from 2008 August 04 15:43:36 (UTC) to 2019 April 03 06:01:37 (UTC)) of \Fermi-LAT Pass 8 data in the energy range 0.3-300~GeV using the software {\it Fermitools}\footnote{https://fermi.gsfc.nasa.gov/ssc/data/analysis/software/}. The region of interest (ROI) is  a $15\arcdeg\times 15\arcdeg$ square in the equatorial coordinate system (J2000, which is adopted throughout the text)
centered at the position of the W30 complex. 
We only select events within a maximum zenith angle of 90$^{\circ}$ to filter out the background \grays from the Earth's limb and apply the recommended filter string  ``($\rm DATA\_QUAL>0) \&\& (LAT\_CONFIG==1$)" in \textit{gtmktime} to choose the good time intervals. The instrument response functions (IRFs) are ``P8R3\_SOURCE\_V2\_v1" for SOURCE events (evclass=128, evtype=3) and ``P8R3\_CLEAN\_V2\_PSF3\_v1" for PSF3 events (evclass=256, evtype=32). The source list is  based on the newly released the fourth \Fermi-LAT source catalog (4FGL; \citealt{4fglv1}) and generated by {\it make4FGLxml.py}\footnote{https://fermi.gsfc.nasa.gov/ssc/data/analysis/user/make4FGLxml.py}. It consists of all spectral and spatial parameters of the 4FGL sources whose positions are within  a radius of 25$^\circ$ centered at W30 as well as the Galactic  diffuse background emission (\emph{gll\_iem\_v07.fits}) and isotropic emission (\emph{iso\_P8R3\_SOURCE\_V2\_v1.txt} for SOURCE events or \emph{iso\_P8R3\_CLEAN\_V2\_PSF3\_v1.txt} for PSF3 events). We use \textit{gtlike} to perform the binned likelihood analysis and record the best-fit results until the optimizer NEWMINUIT successfully converged. We only free the spectral parameters of the sources within $5^{\circ}$ from ROI center with the significance $\geq5\sigma$ and the normalization parameters of the two  diffuse  background components while fitting the source models. Additional adjustment to the model files in the following analyses will be noted in the text.

\subsection{Spatial Analysis}
\label{subsec:spatial}

First, we perform binned likelihood analysis in three energy ranges---1--3~GeV, 3--30~GeV, and 30--300~GeV (hereafter referred to as the low, medium, and high energy range respectively)---for  further research into the energy-dependent behavior of the \gray\ emission around W30 \citep{Ackermann201710GeV}. In the low energy range, only the PSF3 events, of which the quality of the reconstructed direction is the best, are selected to reduce uncertainties caused by large PSF. In the medium or high energy range, the SOURCE events are collected in order to have sufficient counts for the statistic analysis. After the binned likelihood analyses, we subtract the 4FGL sources that are located within radius 1$^\circ$ of the ROI center and not clearly associated with other objects. We notice that 4FGL indicates that 4FGL J1806.2$-$2126 is possibly associated with pulsar PSR J1806$-$2125 or W30 but does not firmly identify it. However, PSR J1806$-$2125 whose $\dot{E}\sim4.3\times10^{34}$\,erg\,s$^{-1}$ and distance is $\sim4.9$\,kpc \citep{atnfpsrcat}, seems incapable of powering such  a bright \gray source that has a flux of $\sim5.0\times10^{-11}$ erg\,cm$^{-2}$\,s$^{-1}$. Thus we subtract  W30 (4FGL J1805.6$-$2136e), \hess (4FGL J1808.2$-$2028e), and 4FGL J1806.2$-$2126 from the best-fit background models. Furthermore, all the parameters except for the normalization parameter of the Galactic  diffuse  background component are fixed to their best-fit values. Finally, we generate the residual test-statistic (TS) maps in the three energy ranges using these adjusted background models.

\begin{figure}
\centering
	\includegraphics[width=0.47\textwidth]{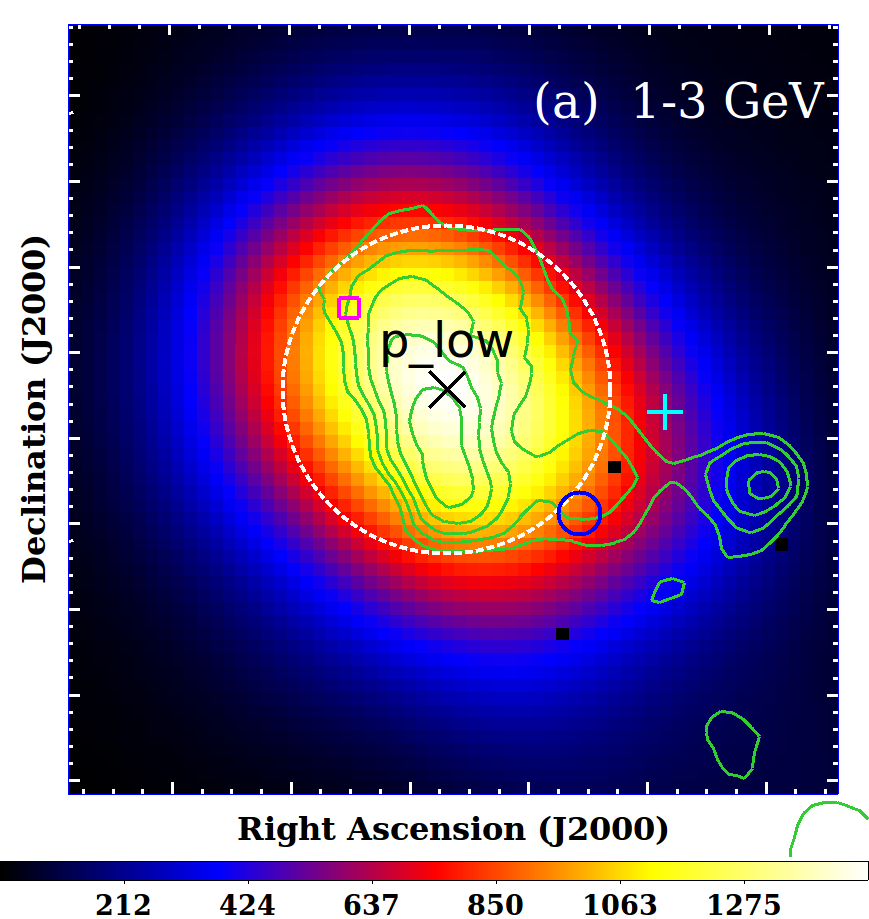}
	\includegraphics[width=0.47\textwidth]{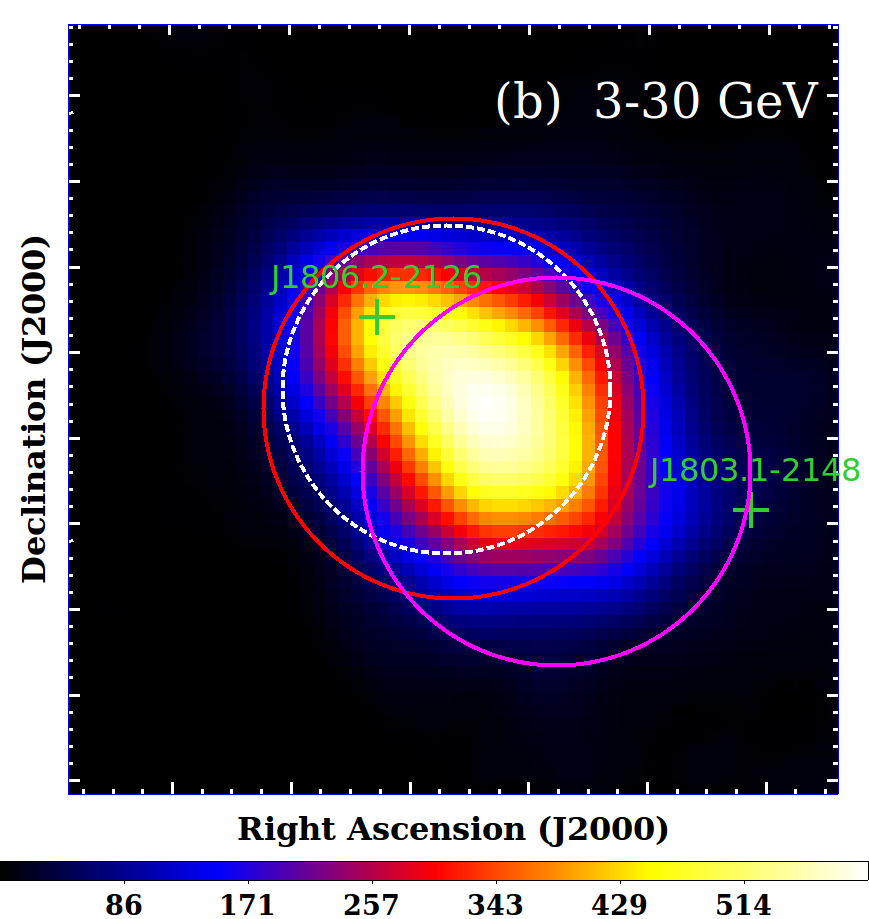}
	\includegraphics[width=0.47\textwidth]{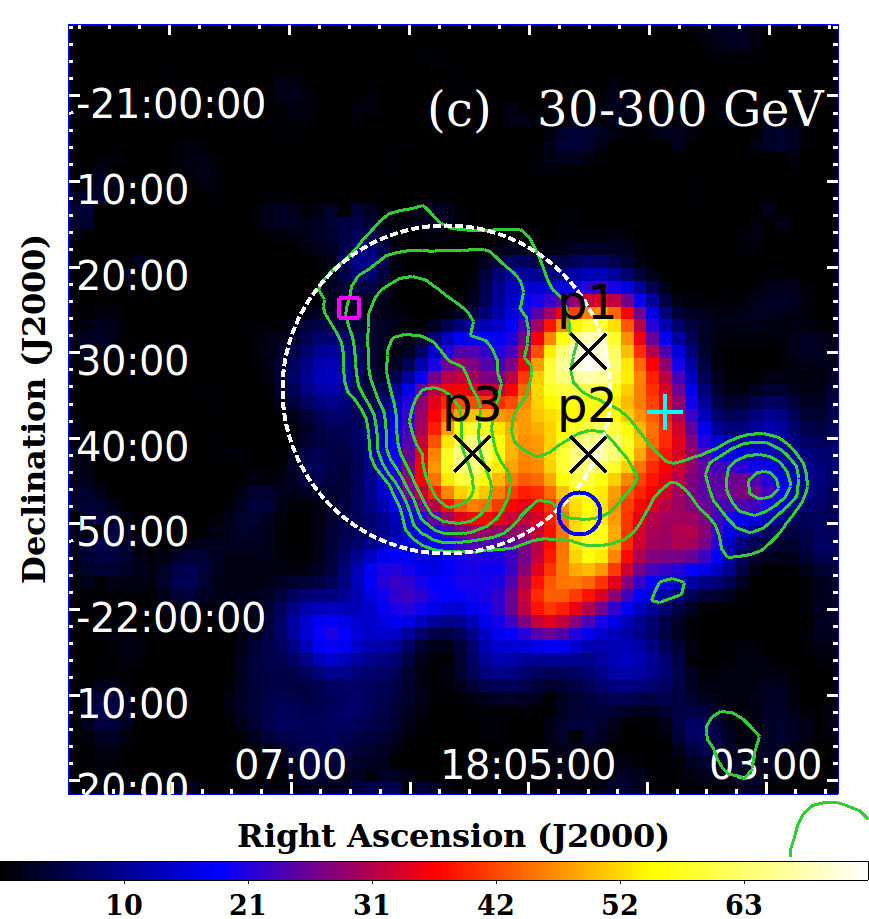}
	\includegraphics[width=0.47\textwidth]{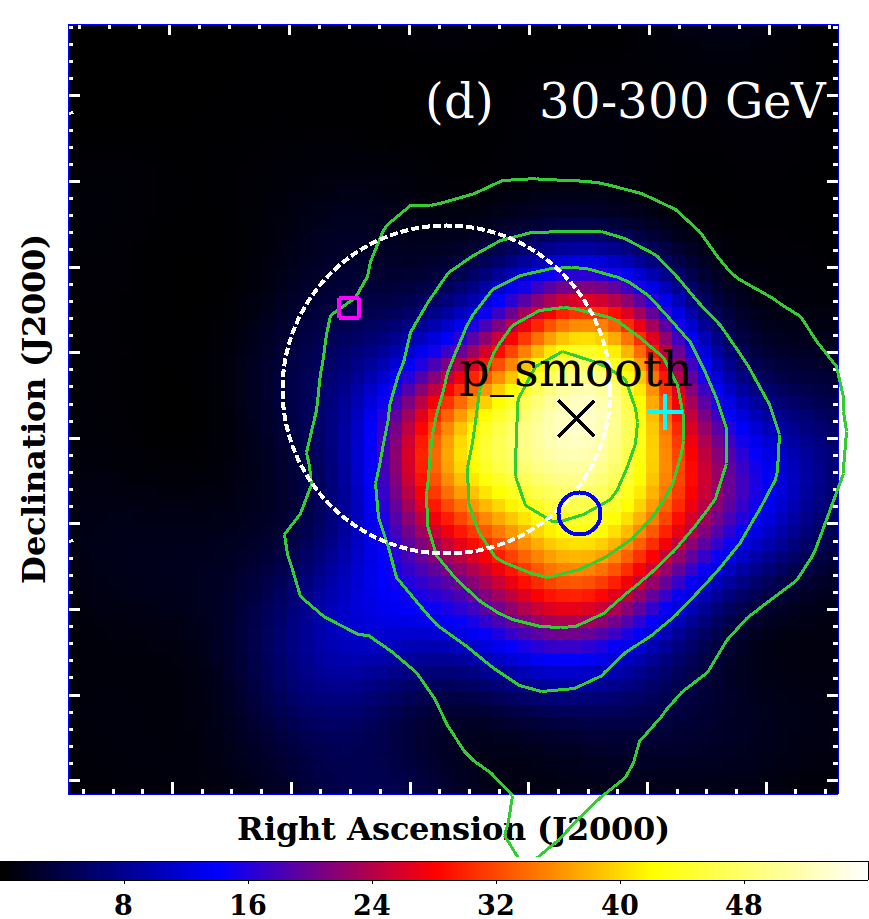}
	\caption{	Background-subtracted TS maps of the $1.5^{\circ}\times1.5^{\circ}$ region
		centered at W30 complex. 
		The image scale of the maps is $0\fdg025$ pixel$^{-1}$, and only (d) is smoothed by a Gaussian kernel of $\sigma=0.2^{\circ}$.
		The dashed white circle is the best-fit spatial template of W30. The blue circle depicts the approximate radio boundary of G8.30$-$0.0. The cyan plus represents the location of PSR\,J1803$-$2137 and the magenta box indicates the location of the OH (1720 MHz) maser detected by \citet{Hewitt2009}. The black crosses show the positions of those TS peaks in different energy ranges. The green pluses indicate the positions of the 4FGL point sources that are within $1\degr$ of the center of ROI. The red circle and magenta circle indicate the spatial templates of W30 and \hess in 4FGL  respectively. Green contours in (a) and (c) show the image of the Galactic plane ``A" (GPA) survey at 8.35\,GHz \citep{Langston2000gpa} at 4.00, 4.75, 5.50, 6.25, and 7.00\,$\rm {Jy\,beam^{-1}}$, meanwhile; green contours in (d) are the significance maps of \hess from H.E.S.S. Galactic Plane Survey at 5, 10, 15, 20, and 25\,$\sigma$.  See Section~\ref{subsec:spatial} for more details.
		}				 
	\label{fig:tsmaps}
\end{figure}

The TS maps in Figure~\ref{fig:tsmaps} show that the spatial distribution of \grays in the W30 region does vary with photon energy. As the photon energies go higher, the emission centroid shifts from northeast to southwest, and the \gray morphology changes from smooth diffusion to irregular distribution. Aiming to disentangle the confusion around W30, we apply specific likelihood-ratio tests on different \gray distribution hypotheses in both low and high energy ranges. The significance of each hypothesis is evaluated by the test statistic TS$=2\log({\mathcal L}_{\rm1}/{\mathcal L}_{\rm0}$, in which ${\mathcal L}_{\rm0}$ is the likelihood of the null hypothesis and ${\mathcal L}_{\rm1}$ is the likelihood of the hypothesis being tested. The statistical significance $\sigma$  can be approximated by  Wilks' theorem \citep{wilks1938}, which equals $\sqrt{TS}$. Following the definition and method in~\citet{Lande2012extend}, we test the extension significance of \gray emission by comparing the likelihood of a uniform disk hypothesis (${\mathcal L}_{\rm disk}$) with that of a point-like source hypothesis (${\mathcal L}_{\rm point}$). The \gray source is considered to be significantly extended only if its ${\rm TS}_{\rm ext}$\,$[{\equiv}2\log({\mathcal L}_{\rm disk}/{\mathcal L}_{\rm point})$) is $\geq16$. The tested radius ($\sigma_{\rm disk}$) of the uniform disk template varies from $0\fdg20$ to $0\fdg50$ with a step of $0\fdg01$. In the above tests, we only free the normalization parameters of the 4FGL sources within $5^{\circ}$ of the ROI center with significance $\geq5\sigma$ and the normalization parameters of the two  diffuse  background components. Besides, the spectra of any newly added sources are modeled with a power law (PL) with the index and normalization parameters left free.

\begin{figure}
	\centering
	\includegraphics[width=0.8\textwidth]{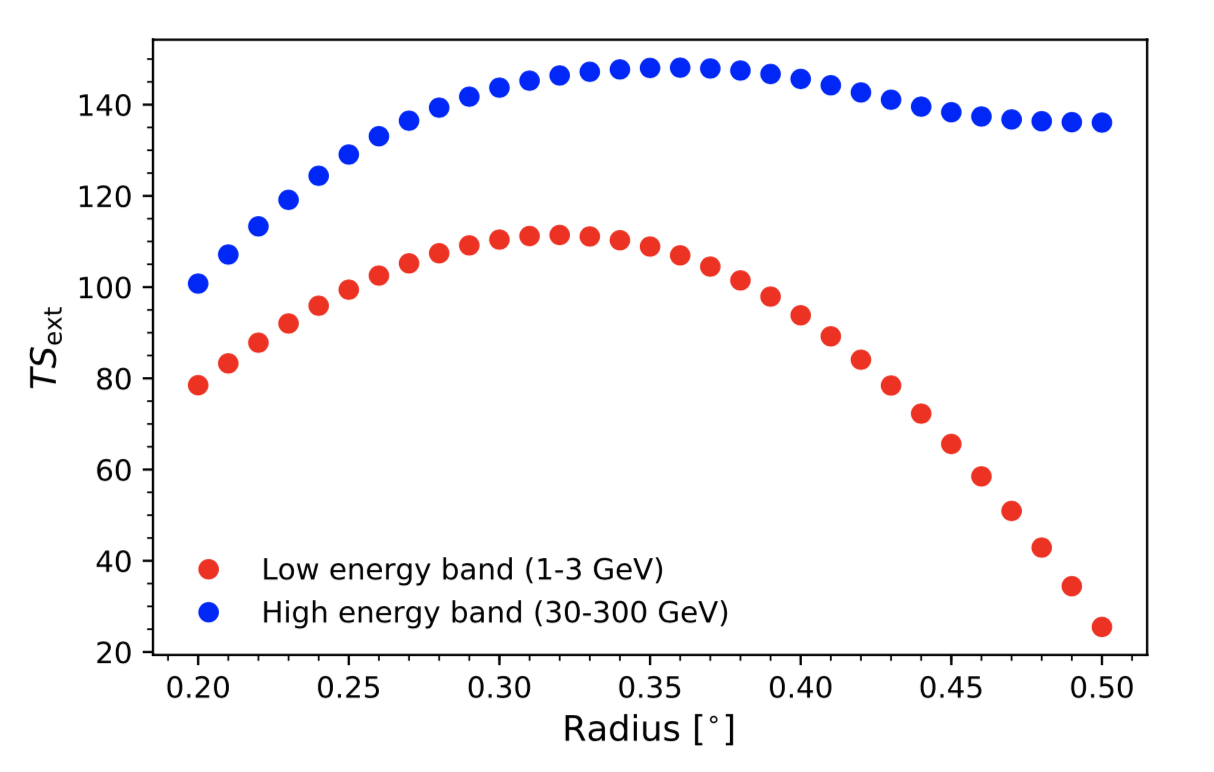}
	\caption{
	Likelihood-ratio test results for uniform disk templates in the spatial analysis.
	 See Section~\ref{subsec:spatial} for the definition of ${\rm TS}_{\rm ext}$ and details.
	 }		 
	\label{fig:ts-ext}
\end{figure}

As can be seen in Figure \ref{fig:tsmaps}, the residual \grays with energy below 3~GeV mostly overlap with \snr and only show one peak TS pixel ($p_{low}$), however, the \grays in high energy range have three close peak TS pixels, $p1$, $p2$ and $p3$, which may be caused by statistical fluctuation or the exits of several point-like sources. On the one hand, we place the point-like source and the disk center at $p_{\rm low}$ to perform the extension test of the low energy range and set the position of the point-like source and disk center at the peak TS pixel $p_{\rm smooth}$ in the smoothed TS map for the extension test of the high energy range. On the other hand, the combinations of two or three point-like sources located at $p1$, $p2$ and $p3$ are checked for the possible existence of several sources in the high energy range. Moreover, we test the spatial templates derived from the \hess significance map measured by H.E.S.S. \citep{Aharonian2006hess} to evaluate the correlation between the GeV \gray emission and the TeV \gray emission. Finally, we compare  the  likelihoods of the  above templates with the spatial model of W30 and \hess from 4FGL to see whether or not there are notable improvements . We list the coordinates and TS values of these peak pixels in Table~\ref{tab:peaks} and depict the extension test results in the low and high energy ranges in Figure~\ref{fig:ts-ext}. The TS values of the best-fit uniform disk templates and other hypotheses are listed in Table~\ref{tab:ext}. In short, in the low energy range, the best-fit spatial model is the uniform disk template with $\sigma_{\rm disk}=0.32\degr\pm0.03\degr$\footnote{The $1\sigma$ uncertainties are determined at where the ${\rm TS}_{\rm ext}$ is lower than the maximum by 1 according to the $\chi^{2}$ distribution.} centered at $p_{low}$, of which the ${\rm TS}_{\rm ext}\approx111.4$ corresponding to a significance of $\sim 10.6\sigma$; meanwhile, in the high energy range, none of the tested hypotheses show a considerable improvement compared to  the \hess template from 4FGL.\footnote{A uniform disk with a radius of $0.377889\degr$ located at $\RA{18}{04}{46}$,$\decl{-21}{44}{06}$} However, it is worth mentioning that the likelihood ratio for HESS significance image of \hess is comparable with the \hess template from 4FGL and better than the best-fit uniform disk template, which implies a morphological correlation between the high energy GeV \grays and TeV \grays around \hesst. Last but not least, the likelihood of the source model does not increase notably when 4FGL source \hess is added in the low energy range or W30 added in the high energy range, which means the contribution from \hess is negligible in the low energy range and W30 is dim in the high energy range. Due to the spatial correlation, we argue that the GeV emission of \hess has the same origin of the TeV emission. However, we cannot rule out that they are produced by different sources, especially given that the multiwavelength studies of the region are incomplete.

\subsection{Spectral Analysis}
\label{subsec:spectra}

We choose the SOURCE events (evclass=128) with energy between 0.3\,GeV and 300\,GeV to perform spectral analysis for a comprehensive understanding of the \gray emission toward the W30 complex. Next, we modify the source model according to the results from the spatial analysis, which includes the deletion of source 4FGL~J1806.2$-$2126 and the replacement of the spatial template of W30 from 4FGL by the uniform disk template (radius$=0.32\degr$) centered at $p_{low}$. As for \hesst, we keep the spatial model from 4FGL. Next, we fit the whole data in the 0.3--300~GeV energy range, setting the spectral type of W30 and \hess to be PL as the null hypothesis. Then, we change the spectral type of W30 to LogParabola (LogP), BrokenPowerLaw (BPL), and PLSuperExpCutoff (PLEC), respectively, to find  which formula fits the data best; meanwhile, the spectrum type of \hess remains to be PL. After this, we keep the best choice for W30 and change the spectral type of \hess to  LogP, BPL, and PLEC, respectively, to find the best spectral formula for \hesst.
The formulae of these spectra are presented in Table~\ref{tab:form}. The spectral type is favored if it has the largest ${\rm TS}_{\rm model}$ defined as $-2\log({\mathcal L_{\rm PL}}/{\mathcal L})$. As shown in Table~\ref{tab:spec_ts}, the spectrum of W30 prefers a BPL distribution and the spectrum of \hess is better presented by LogP. The best-fit spectral parameters of W30 and \hess are listed in Table~\ref{tab:spec_results}. At last, by applying the best-fit spatial and spectral models in the 0.3--300~GeV energy range, we obtain a detection significance of $\sim56\sigma$ and a luminosity of $\sim3.0\times 10^{35}\ds^2$ erg\,s$^{-1}$ for W30, where the $\ds=d/4.5$\,kpc is the distance to \snr in units of a reference value 4.5\,kpc; and for \hesst, we get a detection significance of $19\sigma$ and a luminosity of $\sim1.4\times 10^{35}\du^2$ erg\,s$^{-1}$.

\begin{figure}
	\centering
	\includegraphics[width=0.90\textwidth]{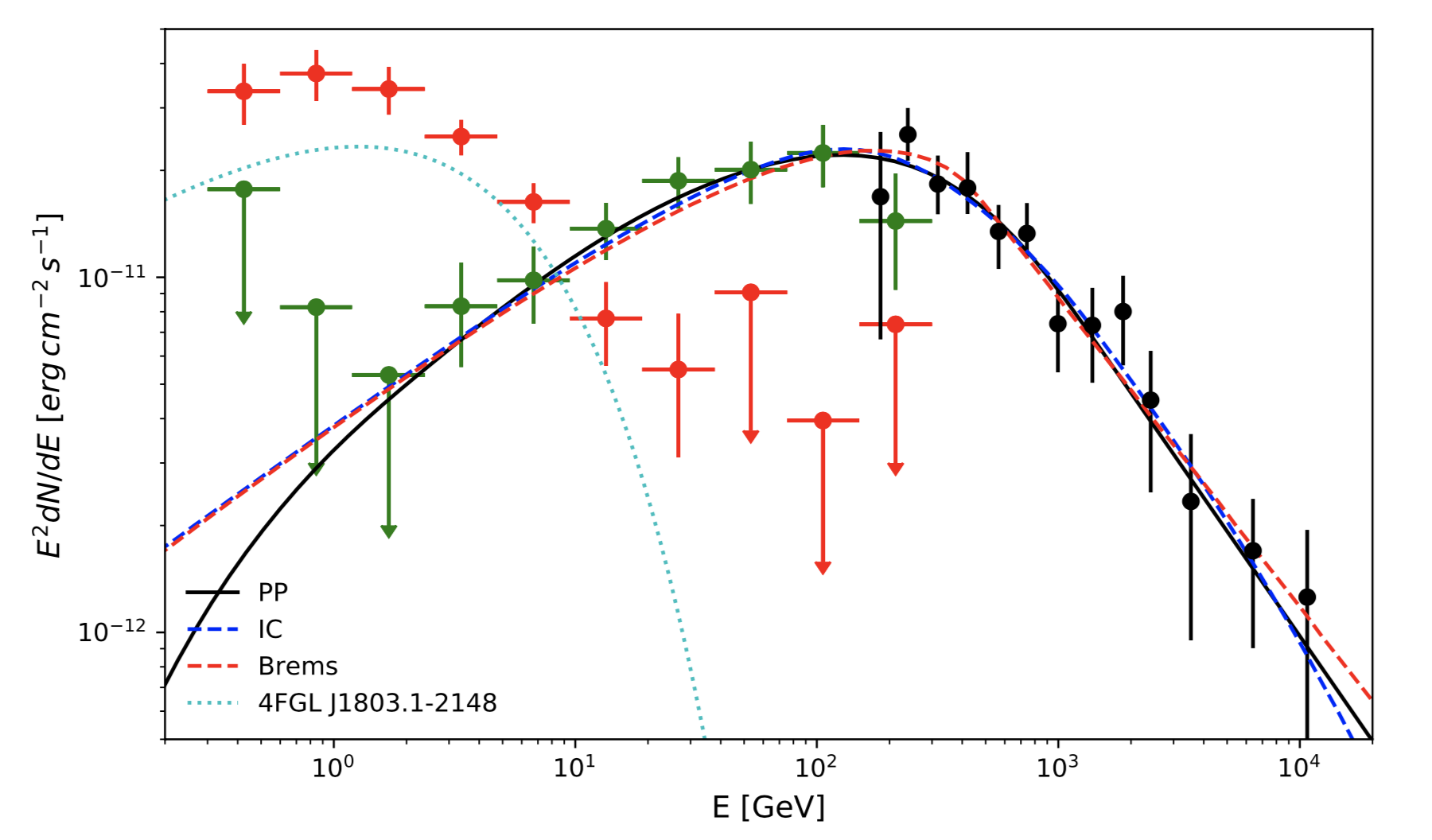}
	\caption{Spectral energy distribution of W30 and \hesst.
    The red and green circles with error bars illustrate the spectra of W30 and \hess obtained from fitting LAT data, respectively.
	Each error bar is a combination of statistical and systematic errors for each energy bin. The arrows show the $95\%$ upper limits. 
	The black circles are the TeV measurements of \hess in \citet{Aharonian2006hess}.
	The dashed blue line depicts the  {\it IC} model, while the dashed red line represents the {\it Brems} model and the solid line shows the {\it pp} model. The parent particle distribution function of all these models is BPL. The cyan dotted line represents the \gray emission from  source 4FGL J1803.1$-$2148. Details of the models are described in Section~\ref{subsec:sed} and the parameters are listed in Table~\ref{tab:fitsed_hess}. 
	} 
	\label{fig:hess_sed}
\end{figure}

The spectral energy distributions (SEDs) of W30 and \hess were extracted from the maximum likelihood analysis of the SOURCE events in 10 logarithmically spaced energy bins within 0.3--300 GeV. During the fitting process, the free parameters only include the normalization parameters of the sources with the significance $\geq5\sigma$ within $5^{\circ}$ from ROI center as well as the Galactic and isotropic  diffuse  background components, 
while all the other parameters are fixed to their best-fit values from the above analysis in the whole energy (0.3--300~GeV) range. In addition, following the method from \citet{Abdo2009W51C}, we estimate the uncertainty caused by the imperfection model of the Galactic  diffuse  background by artificially varying its normalization by $\pm6\%$ from the best-fit value of each energy bin, and record the maximum flux deviation of the source due to above changes as the systematic error. In those energy bins where the TS value of W30 or \hess is smaller than 16, we calculate the $95\%$ upper limit of its flux. Although the first data bin of \hess has a flux with TS value 47.1, we calculate the  upper limit for the SED modeling, concerning the potential influence from the \gray pulsar 4FGL J1803.1$-$2148, which is overlapped by \hess along the line of sight and very bright from 0.3 to 10~GeV. The SEDs of W30 and \hess in GeV together with the H.E.S.S. measurements of \hess in TeV are shown in Figure~\ref{fig:hess_sed}.

As shown in Figure~\ref{fig:hess_sed}, the spectrum of W30 is softer and far below the H.E.S.S. measurements at higher energies and the LAT spectrum of \hess 
is smoothly connected to the H.E.S.S measurements. Such spectral discrepancy between  W30 and \hess is consistent with their energy-dependent morphological behavior, both reveal the domination of W30 in the lower energy range ($\sim0.3-10$~GeV) and very likely correlation between GeV observation and TeV measurement on \hess.

\section{Discussion}
\label{sec:dis}

The above studies on the spatial and spectral properties of the  diffuse  \grays around W30 complex reveal their strong energy-dependent behavior. As a result, we distinguish them into two separate extended components: one of them, namely W30, is mostly overlapped with the \snr and compatible with the  size  of the SNR; the second one, namely \hesst, shows both spatial and spectral consistency with the TeV measurement. So we will involve the radio emission of the entire \snr region \citep{Kassim1992vla} in the SED modeling of \gray emission from W30 and fit the LAT and H.E.S.S. data together to model the \grays from \hesst.

\subsection{Interpreting the SEDs of W30 and \hesst}
\label{subsec:sed}

\begin{figure}
	\centering
	\includegraphics[width=0.90\textwidth]{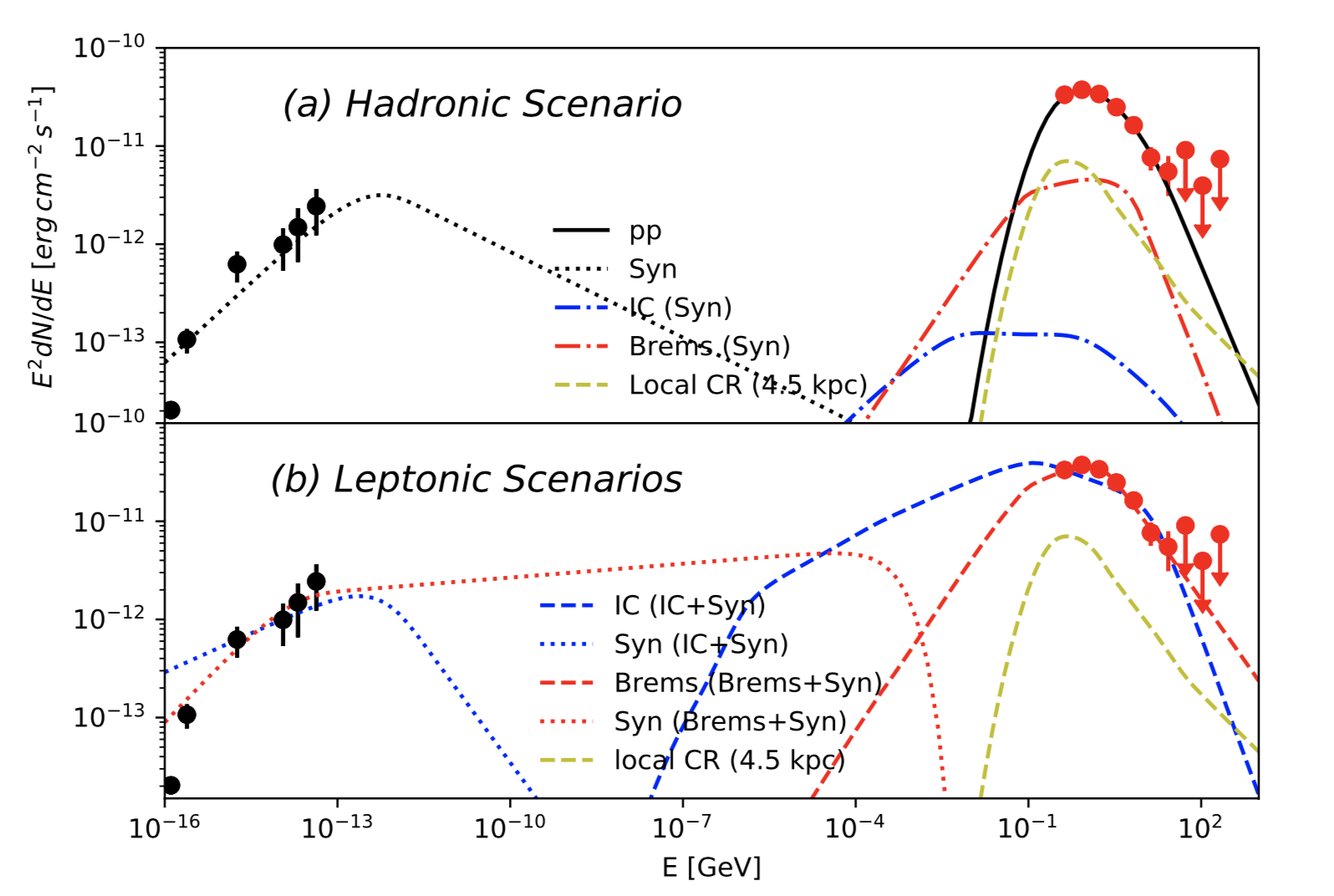}
	\caption{Multiwavelength spectrum of W30.
	The red circles with error bars show the spectrum of W30 obtained from the spectral analysis in Section~\ref{subsec:spectra}. 
	The black circles represent the radio emission from the entire region of \snrt \citep{Kassim1992vla}.
	In panel (a), the solid line represents the best-fit {\it pp} model
	and the dotted line shows the synchrotron radiation in order to model the radio data.  
	The dotted-dashed blue line and red line illustrate the corresponding IC scattering and bremsstrahlung emission generated by the same electron population.
	In panel (b), the blue lines and red lines depict the best-fit {\it IC+Syn} model and {\it Brems+Syn} model of \gray and radio emission from W30 region respectively.  
    The parent particle distribution function of all the models is BPL.
    The yellow line represents \grays produced by the local CRs assuming the total MC mass of the W30 region is $3.1\times$ 10$^{5}$ $M_{\odot}$ at the distance of 4.5\,kpc.
	Details are described in Section~\ref{subsec:sed} and the parameters are listed in Table~\ref{tab:fitsed_w30}.
	} 
	\label{fig:w30_sed}
\end{figure}

To understand the origin of \grays from W30 and \hesst, we use the {\it Naima} package \citep{Zabalza2015naima} for modeling the SEDs presented in Figure~\ref{fig:w30_sed} and Figure~\ref{fig:hess_sed}. {\it Naima} allows Markov Chain Monte Carlo (MCMC) fitting using {\tt emcee} \citep{Foreman2013emcee} while computing nonthermal radiation from relativistic particle populations. The likelihood function (${\mathcal L}$) of a given model can be related to the $\chi^2$ parameter as $\chi^2=-2\ln{\mathcal L}$, so the maximization of the log-likelihood is equivalent to a minimization of $\chi^2$ \footnote{https://naima.readthedocs.io/en/latest/mcmc.html}.

In the hadronic scenario, we assume that the \grays mainly originate from the decay of $\pi^0$ mesons produced by the  interactions of relativistic protons with the ambient gas ({\it pp} model).  Besides, we fit the radio data of \snr independently with synchrotron radiation ({\it Syn} model), then calculate the corresponding inverse Compton (IC) and nonthermal bremsstrahlung radiation from the same electron population to compare their possible contribution to that of the $pp$ interactions.  In principle the magnetic field cannot be derived from a hadronic model, but as in \citet{Ajello2012G8.7} we also assume $B=100\,\rm \mu G$, which can be caused by the compression of gas in the MCs near the SNR shocks. In the leptonic scenario, we assume the radio emission from the entire region of \snr \citep{Kassim1992vla} and \grays from W30 are generated by the same electron population. Their interaction with the magnetic field, i.e., the synchrotron radiation, accounts for the radio emission; meanwhile, the IC scattering of low energy photons ({\it IC+Syn} model) or nonthermal bremsstrahlung radiation via interacting with thermal particles  ({\it Brems+Syn} model). Note that we exclude the first two data points of the radio spectrum during the whole fitting process, in the consideration of the low-frequency turnover caused by the the absorption from the foreground \ion{H}{2} regions \citep{Kassim1992vla}.
As for GeV and TeV \gray emission from \hesst, we try to reproduce the spectrum assuming these \grays are generated via relativistic electrons  IC scattering of soft photons (IC model) or nonthermal bremsstrahlung radiation ({\it Brem} model). We also test the {\it pp} model to see whether a hadronic origin is possible or not.

For the $pp$ interaction, we use the cross-sections from \citet{Kafexhiu2014pp} and set the value of $n_{\rm H}$, the average number density of the target proton, to be 100\,${\rm cm^{-3}}$, which is a reasonable value given the total molecular mass in the W30 region ($ M \sim2.1-3.1\times$ 10$^{5}$ $M_{\odot}$, \citet{Takeuchi2010nanten}). We use  the parameterization in \citet{Aharonian2010sync} to calculate the synchrotron radiation. To model the bremsstrahlung radiation, we use the approximation from \citet{Baring1999} assuming the total ion number density $n_0$ is 100\,${\rm cm^{-3}}$. The formulae from \citet{Khangulyan2014ic} are applied for the calculation of IC radiation. The seed photon field for relativistic electrons to scatter includes the Cosmic Microwave Background, far infrared emission from dust, and starlight. The later two components are adopted from the local interstellar radiation field calculated by \citet{Popescu2017radif}. The parent particle distribution functions in energy, which include PL, Exponential-Cutoff PowerLaw (ECPL),  
and BPL together with the corresponding free parameters in the fitting process are listed in Table~\ref{tab:form}. The required total energies in electrons and protons, $W_e$ and $W_p$, are calculated for particles with energy above 1~GeV. The SED modeling results of W30 and \hess are listed in Table~\ref{tab:fitsed_w30} and Table~\ref{tab:fitsed_hess}, respectively. And the best-fit results of different models are illustrated in Figure~\ref{fig:w30_sed} and Figure~\ref{fig:hess_sed}.

\subsection{Origin of \gray Emission from W30 and \hesst}

Figure~\ref{fig:w30_sed} demonstrates a suppression in low energy  in the SED of W30, although it is not significant enough to claim a pion-bump feature. In the middle-aged SNRs interacting with MCs, $pp$ interactions are indeed regarded as the main mechanism of \gray emission. They already have been seen in other SNRs interacting with MCs such as W44, IC\,433, and Kes\,41. All of them show soft GeV spectra with PL photon indices $\ge 2$ and the 1--100~GeV luminosities in the order of $10^{35}$\,erg\,s$^{-1}$ \citep[e.g., ][Table~3 therein]{Abdo2010W44SCi,Abdo2010IC443,Ackermann2013,liub2015}. We find that the BPL type proton distribution fits the \gray data best. In the BPL fitting, 
the break energy is  $\sim34$~GeV above which the proton index is softened from about $2.2$ to $3.5$. The derived proton spectrum is very similar to that of W44 \citep{Ackermann2013}. To explain the similar energy break in W44, \citet{Malkov2011} argued that, in a dense environment near the interacting SNR, strong ion--neutral collisions in an adjacent MC lead to Alfv\'{e}n wave evanescence, which introduces fractional particle losses and results in the steepening of the energy spectrum of accelerated particles by exactly one power. The bright OH (1720 MHz) maser detected along the eastern edge of W30 \citep{Hewitt2009} reveals that  SNR \snr is interacting with nearby MCs. Thus the same mechanism  may apply here and provide a satisfactory explanation of the energy break. The total energy required proton energy is estimated to be  $\sim 3\times10^{49}$\,erg assuming for the target proton density of $100\,\rm cm^{-3}$  at a distance of 4.5\,kpc. 

 On the other hand, we cannot rule out the IC or bremsstrahlung origin of the \gray emissions. In both cases the BPL distribution of relativistic electrons is favored. However, for the {\it IC} model the derived energy budget for the relativistic electrons is as high as $\sim 6\times10^{49}$\,erg, which is almost 10\% of the typical kinetic energy of a supernova explosion ($\sim10^{51}$\,erg).  For bremsstrahlung origin a relatively large ratio of electrons to protons, $K_{ep}\sim0.1$, is required.  This is much larger than the fiducial value $0.01$ predicted by the diffusive shock acceleration theory \citep{Bell1978A} and observed at Earth.  Although this possibility cannot be excluded \citep[see, e.g., ][]{merten17}.

It should be noted that there is another possible particle accelerator located at the western boundary of W30, G8.30$-$0.0. It is a shell-like small size ($5'\times4'$) SNR initially identified in 20\,cm emission by \citet{Helfand2006}. Recently, \citet{Kilpatrick2016} reported the detection of broad molecular line regions along the western boundary of this SNR, suggesting interactions between G8.30$-$0.0 and the MC at a systematic velocity near $+2.6\,\kms$ at a distance of $\sim16$\,kpc, which is consistent with the distance obtained from the brightness-to-diameter relation. Assuming that  the \grays of W30 are emitted from a distance of 16\,kpc and the target proton density is $100\,\rm cm^{-3}$, the energy budget for the parent protons of $pp$ interaction  is $\sim (0.5-0.8)\times10^{51}$\,erg, which is too high for an ordinary SNR. This makes the association between G8.30$-$0.0 and the \gray\ emission unlikely.

It has been proposed in the paper of \citet{hess2018pwnpop}, that \hess is a PWN powered by the pulsar \psrt.  The \Fermi-LAT data extend the spectrum to lower energy. From the fitted results we found a broken PL type spectrum of electrons under the assumption that the GeV--TeV \gray emission is due to the IC scattering. Taking into account the age of about 16\,kyr and the break energy of about 1 TeV, the magnetic field should be close to $15\,\rm \mu G$.  The {\it Brems} model and the {\it pp} model could also fit the spectral data, but both require a very dense environment ($n_0=100\,\rm cm^{-3}$) and extremely hard spectrum (with an index of about 1.5) for the parent particles, which makes them quite unlikely.

\section{Summary}
\label{sec:sum}
In this paper we present a detailed analysis based on about 10\,yr data of \Fermi-LAT observations of the W30 region. We resolve the emission in this region into two extended \gray sources. One source reveals a soft  \gray spectrum with a pion-bump-like spectral feature. This is similar to the spectra of other middle-aged SNRs interacting with MCs. We attribute it to the SNR W30 (\snrt) itself. The parent proton spectrum is best fitted as a broken PL with a break at about 30\,GeV, which can be explained by the fractional particle loss due to Alfv\'{e}n wave evanescence in the dense molecular region \citep{Malkov2011}. The second source shows a significantly harder spectrum, and coincides with the H.E.S.S. source \hess not only in the spatial distribution but also in the spectral distribution. The GeV--TeV emission of \hess can be explained naturally by the one-zone model of cooling dominated PWN, which is most likely illuminated by \psrt. We note that the W30 complex is also an active star-forming region.  Such kind of systems are already identified as potential efficient CR accelerators \cite[see, e.g,,][]{aharonian19, fermi_cygnus, katsuka17} and all reveal extended \gray emission with hard spectrum. The derived \gray emission from W30, however, has a  much softer spectrum. Thus, we attribute the GeV emission to the SNR itself. However, we cannot rule out the high energy emission of  \hess may partly come from the CRs accelerated in the star-forming region.

The overlap of \gray sources could be common for the galactic disk, due to the extended nature of sources and the limited angular resolution of telescopes, especially in crowded regions as the  star-forming regions. Thus additional care must be taken in  explaining the origin of  \gray emission in such regions. The forthcoming Cerenkov Telescope Array (CTA; \citealt{Actis2011},  with a larger field of view, better angular and energy resolutions, and better sensitivity compared to \Fermi-LAT and current Cerenkov telescopes,  should be able to effectively separate these accidentally overlapping  diffuse  structures. And, of course, a thorough multiwavelength study  of such regions would be important to pin down their nature.

\begin{acknowledgements}

B. L. acknowledges the financial support of the China Scholar Council (No.201706190129). This research is also supported by the 973 Program under grants 2017YFA0402600 and 2015CB857100 and the NSFC under grants 11773014, 11633007, and 11851305.
	
\end{acknowledgements}

\bibliographystyle{aasjournal}
\bibliography{cite-w30}
\bigskip
\bigskip

\begin{deluxetable}{c|ccc}
	\tablecaption{The likelihood-ratio test results of the spatial analysis}
	\tablewidth{0pt}
	\startdata
	\hline
	\hline
	Energy Range & Spatial Model  & ${\rm TS}_{\rm model}$\tablenotemark{a} & +DoF\tablenotemark{b}\\ 
	\hline
	\multirow{5}{*}{1--3~GeV} 	& 1 point-like source\tablenotemark{c}  & 0  &  0 \\
	& 	Uniform disk\tablenotemark{d} & 111.4 & 0\\
	& 	Uniform disk + \hesst\tablenotemark{\star} &  111.8 & -2\\
	& 	W30\tablenotemark{\star}  & 97.7   & 0\\
	%& 	1 point source + \hesst\tablenotemark{\star}  &       & -2 \\  
	& 	W30\tablenotemark{\star} + \hesst\tablenotemark{\star}  & 97.7 & -2\\
	\hline
	\multirow{8}{*}{30--300~GeV}	& 1 point-like source\tablenotemark{e} & 0   & 0 \\
	&   1 point-like sources ($p1$, $p2$, $p3$) & 18.5, 6.4, 7.0 & 0 \\
	&	2 point-like sources ($p1$+$p2$, $p1$+$p3$, $p2$+$p3$) & 55.9, 67.9, 52.0& -2\\ 
	&	3 point-like sources ($p1$+$p2$+$p3$)  & 98.8 &  -4  \\  
	&	Uniform disk\tablenotemark{f} & 148.1 & 0  \\
	&	H.E.S.S significance map\tablenotemark{g} &176.3,174.4 & 0\\	
	&	\hesst\tablenotemark{\star} &179.5  & 0 \\
	&  W30\tablenotemark{\star}+\hesst \tablenotemark{\star} & 182.5 & -2\\	
	\hline
	\enddata
	\tablenotetext{a}{ $TS_{\rm model}=-2\log({\mathcal L_0}/{\mathcal L})$. One point-like source as null hypothesis.}
	\tablenotetext{b}{ Additional degrees of freedom}
	\tablenotetext{c}{ Null hypothesis. Located at $p_{\rm low}$.}
	\tablenotetext{d}{ Best-fit uniform disk template (radius$=0.32\degr$).}
    \tablenotetext{e}{ Null hypothesis.  Located at $p_{\rm smooth}$.}
	\tablenotetext{f}{ Best-fit uniform disk template (radius$=0.36\degr$).}
	\tablenotetext{g}{The templates are created from \hess significance map from H.E.S.S. Galactic Plane Survey, one for $\sigma>4$ and the other $\sigma>10$.}
	\tablenotetext{\star}{This mark indicates the spatial template is adopted from 4FGL.}
	\label{tab:ext}
\end{deluxetable}

\begin{deluxetable}{cccc}
 	\tablecaption{Locations and TS of the peak pixels in Figure~\ref{fig:tsmaps} \tablenotemark{a}}
 	\tablewidth{0pt}
 	\startdata
 	\hline
 	\hline
 	Name &  Position  &TS\\ 
 	\hline
 	$p_{\rm low}$&$\RA{18}{05}{41}$, $\decl{-21}{34}{30}$ & 1486.5\\
 	$p1$&$\RA{18}{04}{30}$, $\decl{-21}{30}{00}$ & 72.9\\
 	$p2$&$\RA{18}{04}{30}$, $\decl{-21}{42}{00}$ & 66.1 \\
 	$p3$&$\RA{18}{05}{28}$, $\decl{-21}{42}{00}$ & 65.8\\
 	$p_{\rm smooth}$ &$\RA{18}{04}{29}$, $\decl{-21}{31}{30}$ & 53.2\\
 	\enddata
 	\tablenotetext{a}{Equatorial coordinate system. See Section\ref{subsec:spatial} for specific definitions.}
 	\label{tab:peaks}
\end{deluxetable}

\begin{deluxetable}{c|c|c|c}
	\tablecaption{Formulae for \gray spectra and parent particle distribution }
	\tablewidth{0pt}
	\startdata
	\hline
	\hline
 &	Name & Formula  &Free parameters   \\%Function
	\hline  
\multirow{4}{*}{\gray}&	PL   & $ \mathrm{d}N/\mathrm{d}E = N_0 (E/E_0)^{-\Gamma} $   & $N_0$, $\Gamma$  \\
&	PLEC & $ \mathrm{d}N/\mathrm{d}E = N_0 (E/E_0)^{-\Gamma} \exp(-E/E_\mathrm{cut})$  &  $N_0$, $\Gamma$, $E_\mathrm{cut}$\\
&	LogP & $ \mathrm{d}N/\mathrm{d}E = N_0 (E/E_\mathrm{b})^{-\Gamma - \beta \log(E/E_\mathrm{b})} $  & $N_0$, $\Gamma$, $\beta$\\
&	BPL &  	$\mathrm{d}N/\mathrm{d}E=\begin{cases} N_0(E/E_\mathrm{b})^{-\Gamma_1} &  \mbox{: } E<E_\mathrm{b} \\ N_0(E/E_\mathrm{b})^{-\Gamma_2} & \mbox{: }E>E_\mathrm{b} \end{cases}$ & $N_0$, $\Gamma_1$, $\Gamma_2$\\
	%PLEC & $ \mathrm{d}N/\mathrm{d}E = N_0 (E/E_0)^{-\Gamma_1} {\exp(-E/E_\mathrm{cut})}^{-\Gamma_2}$  \\
	\hline
\multirow{4}{*}{Particle} &	PL   & $ N(E) = A (E/E_ 0)^{-\alpha} $   & A, $\alpha$  \\
&	ECPL & $N(E) = A (E/E_ 0)^{-\alpha} {\exp(-(E/E_\mathrm{cut})}^{\beta}) $ &A, $\alpha$, $\beta$, $ E_\mathrm{cut}$ \\
	%LogP & $ N(E) = A (E/E_0)^{-\alpha-\beta \log(E/E_0)}  $ & A, $\alpha$, $\beta$  \\
&	BPL &  $N(E) =\begin{cases} A(E/E_0)^{-\alpha_1} & \mbox{: }E<E_\mathrm{b} \\ A(E_\mathrm{b}/E_0)^{(\alpha_2-\alpha_1)}(E/E_0)^{-\alpha_2} & \mbox{: }E>E_\mathrm{b} \end{cases}$&  A, $\alpha_1$, $\alpha_2$, $E_\mathrm{b}$\\
	\enddata
	\label{tab:form}
\end{deluxetable}

\begin{deluxetable}{c|ccccc}
	\tablecaption{ The likelihood-ratio test results (${\rm TS}_{\rm model}$) from the spectral analysis of W30 and \hesst}%0.3-300 GeV 
	\tablewidth{0pt}
	\startdata
	\hline
	\hline
    Spectral Type   & PL & LogP        & BPL           & PLEC\\
    \hline
   W30\tablenotemark{a}   & 0  & 25.8  &26.8    &  23.1  \\
   \hline
  \hess\tablenotemark{b}  & 0  &14.0    & -8.1   & 10.6  \\
      \hline
  	\enddata
%	\tablenotetext{a}{$TS_{\rm model}=-2\log({\mathcal L_{\rm PL}}/{\mathcal L})$}
	\tablenotetext{a}{The spectrum type of \hess  is PL}
	\tablenotetext{b}{The spectrum type of W30 is BPL}
	\label{tab:spec_ts} 
\end{deluxetable}

\begin{deluxetable}{c|ccc|ccc}
	\tablecaption{Best-fit Results of Spectral Analysis  for \gray Emission in 0.3--300~GeV} %0.3-300GeV for W30 and \hesst 
	\tablewidth{0pt}
	\startdata
	\hline
	\hline
 \multirow{2}{*}{ Name (Type) } &\multicolumn{3}{c}{Spectral  Parameters}  &Flux &\multirow{2}{*}{TS} \\ %Luminosity\tablenotemark{a} 
 % \cline{2-4}
    & $\Gamma_1$ or$\Gamma$  & $\Gamma_2$ or $\beta$ &$E_{\rm b}$(GeV) &($10^{-8}$ photons $cm^{2}\,s^{-1}$) &\\
    %	&  & &&  & \\% ($10^{35}$ $erg\,s^{-1}$)
  	\hline
     W30 (BPL)& 2.20$\pm$0.04 & 2.80$\pm$0.09  & 3.00$\pm$0.10  & 7.94$\pm$2.75 & 3135.3\\ %lumi 4kpc 2.34e+35 +/-4.12e+34
     \hline
    \hess (LogP) &0.61$\pm$0.21 & 0.17$\pm$0.03 & 1.00\tablenotemark{a} & 0.37$\pm$0.08 &375.9\\ %lumi 4kpc  1.40e+35+/-5.25e+34
	\hline
	\enddata
	\tablenotetext{a}{$E_b$ is a scale parameter, which should be set  near the lower energy range of the spectrum being fit and is usually fixed, see \citet{Massaro2004}. }
	\label{tab:spec_results} 	
\end{deluxetable}

\begin{deluxetable}{c|ccccccc}
	\tablecaption{W30 SED Fit Results for Different Radiation Models} 
	\tablewidth{0pt}
	\startdata
	\hline
	\hline
	\multirow{3}{*}{Model} & Parent&\multicolumn{4}{c}{Parameters}& \multirow{2}{*}{ $W_p$ or $W_e$}  &  \multirow{2}{*}{MLL \tablenotemark{a}}  \\
	\cline{3-6}
	&Particle&  $\alpha$ or $\alpha_1$  &  $\beta$ or $\alpha_2$ & $E_{\rm cut}$ or $E_{\rm b}$ &B  & & \\ 
	&Distribution &          &        &   (GeV)    & ($\mu$G) &($10^{49}{d_{4.5}}^2$\,erg)\tablenotemark{b} & \\ 
	\hline
	\multirow{4}{*}{{\it pp}}&PL &\remp{2.65}{0.03}{0.04} & - & -  &-   & \remp{4.09}{0.38}{0.41}  &-4.12   \\
	&BPL & \remp{2.16}{0.29}{0.22} & \remp{3.47}{0.48}{0.89} & \remp{34.03}{14.00}{22.44}  &-  &\remp{2.73}{0.49}{0.54} &-0.31  \\ 
	&ECPL & \remp{2.19}{0.31 }{0.21} & 1 &\remp{92.90}{45.08}{104.47}   &- & \remp{2.89}{0.48}{0.54} & -0.50 \\ 
	&ECPL & \remp{1.45}{0.43}{0.38} & \remp{0.54}{0.15}{0.25} &\remp{6.94 }{4.23}{10.37} &- &\remp{2.37}{0.42}{0.54} & -0.45\\  
	\hline
	\multirow{4}{*}{{\it Syn}} &PL &\remp{3.14}{1.29}{1.20}  & - & -  &100 &\remp{0.04}{0.01}{0.08} & -1.70  \\
	&BPL &\remp{1.91}{0.98}{0.59} &\remp{3.58}{1.10}{0.97} & \remp{8.34}{6.23}{12.72}  &100 &\remp{0.03}{0.01}{0.03} &-0.12\\ 
	&ECPL &\remp{2.76}{1.76}{1.46} &1 &\remp{13.34}{9.12}{10.93}  &100  &\remp{0.03}{0.01}{0.01}  &-1.71\\
    &ECPL &\remp{2.37}{1.56}{1.69} &\remp{2.29}{1.53}{2.29} &\remp{3.26}{2.18}{3.60}  &100  &\remp{0.03}{0.01}{0.01} & -1.71\\
	\hline
	&PL &\remp{3.54}{0.05}{0.06}  & -  &-  &\remp{1.99}{0.30}{0.28} &\remp{46.49}{7.80}{10.02}  &   -8.69 \\
	 IC &BPL & \remp{2.47}{0.23}{0.16}  & \remp{4.61}{0.43}{0.28 } & \remp{39.87}{13.99}{51.06}  &\remp{4.48}{1.30}{4.80}  &\remp{6.18}{4.06}{4.29}   &  -1.52  \\ 
	{\it +Syn}&ECPL & \remp{2.10}{0.38}{0.36}  & 1   &\remp{32.90}{12.58}{20.03}  &\remp{4.13}{0.98}{1.57}  &\remp{6.40}{2.46}{3.91}  &   -2.15 \\ 
	&ECPL & \remp{1.95}{0.50}{0.41}  & \remp{1.33}{0.51}{1.03}  &\remp{28.63}{13.63}{16.85}  &\remp{3.86}{0.82}{1.09} &\remp{6.39}{2.09}{3.51}   & -2.11 \\  
	\hline
   &PL &\remp{2.57}{0.03}{0.03}  & - & -  &\remp{35.11}{5.06}{4.59}  &\remp{0.18}{0.01}{0.01}    &  -7.60  \\
	{\it Brems}&BPL &\remp{1.75}{0.60}{0.34} &\remp{2.91}{0.16}{0.31} & \remp{3.40}{1.23}{2.86} &\remp{34.31}{5.74}{5.31}& \remp{0.19}{0.01}{0.01} & -0.77\\
    {\it +Syn}&ECPL & \remp{1.75}{0.34}{0.26}  & 1  &\remp{9.18}{3.31}{5.66}  &\remp{34.85}{5.84}{5.77}  &\remp{0.18}{0.01}{0.01}   &  -1.45\\  
	&ECPL & \remp{1.65}{0.26}{0.23} &\remp{0.56}{0.13}{0.24}   &\remp{3.08}{1.84}{3.82}  &\remp{35.28}{5.60}{5.52}  &\remp{0.18}{0.01}{0.01} & -1.02   \\ 
	\hline
	\enddata
	\tablenotetext{a}{Maximum log-likelihood.}
	\tablenotetext{b}{$d_{4.5}{\equiv}d/(4.5\,{\rm kpc})$}
	\label{tab:fitsed_w30}
\end{deluxetable}

\begin{deluxetable}{c|ccccccc}
	\tablecaption{\hess SED Fit Results for Different Radiation Models}
	\tablewidth{0pt}
	\startdata
	\hline
	\hline
	\multirow{3}{*}{Model}  &Parent &\multicolumn{3}{c}{Parameters}    & \multirow{2}{*}{$W_p$ or $W_e$}  & \multirow{2}{*}{MLL\tablenotemark{a}}  \\
	\cline{3-5}
	&Particle& $\alpha$ or $\alpha_1$ & $\beta$ or $\alpha_2$ & $E_{\rm cut}$ or $E_{\rm b}$ & & \\  
	&Distribution &         &        &   (GeV)        & ($10^{48}{d_4}^2$\,erg)\tablenotemark{b}&\\  
	\hline
	\multirow{4}{*}{{\it pp}}&PL &\remp{2.27}{0.01}{0.01}  & - & -   &\remp{15.2}{1.14}{0.92}   & -68.34  \\
	&BPL & \remp{1.48}{0.15}{0.09} & \remp{3.01}{0.14}{0.21} & \remp{1949.32}{434.33}{612.40} &\remp{7.31}{0.46}{0.45}&-3.54 \\ 
	&ECPL &\remp{1.36}{0.48}{0.22} & 1 &\remp{2920.40}{1494.00}{1543.41} &\remp{6.81}{6.81}{6.13} & -6.07  \\ 
	&ECPL &\remp{1.43}{0.14}{0.14} & \remp{0.73}{0.16}{0.31} &\remp{2759.72}{1463.73}{2531.39}&\remp{7.15}{0.45}{0.48}&-4.63\\  
	\hline
	\multirow{5}{*}{IC}&PL & \remp{2.90}{0.03}{0.02}  & - & -  &\remp{31.72}{4.82}{4.63}    &   -47.88  \\
	&BPL & \remp{ 1.82}{0.36}{0.21} &\remp{3.65}{0.15}{0.21} &\remp{ 1111.72}{258.97}{351.36} &\remp{0.76}{0.25}{0.33} &-3.87\\
	&ECPL & \remp{1.48}{0.75}{0.36} & 1 &\remp{1041.58}{450.39}{505.16}  &\remp{0.52}{0.21}{0.27}  &   -7.54 \\ 
	&ECPL & \remp{1.32}{0.21}{0.20} &\remp{0.54}{0.06}{0.09}  &\remp{231.87}{109.16}{186.72} &\remp{0.65}{0.11}{0.11}&-5.30\\
	\hline
	\multirow{4}{*}{{\it Brems}}&PL &\remp{2.26}{0.03}{0.02}  & - & - &\remp{1.27}{0.17}{0.14}   & -73.08   \\
	&BPL & \remp{1.59}{0.11}{0.11} & \remp{2.94}{0.11}{0.16} & \remp{471.42}{123.52}{167.66}  &\remp{0.99}{0.06}{0.08} &-4.37\\
	&ECPL& \remp{1.62}{0.09}{0.14} & 1  &\remp{1317.68}{315.17}{458.27}    &\remp{0.96}{0.01}{0.11}  &  -9.31\\ 
	&ECPL& \remp{1.263}{0.13}{0.11}& \remp{0.45}{0.06}{0.08} & \remp{102.17}{63.05}{115.27} & \remp{0.97}{0.06}{0.05} &-4.67\\  
	\hline
	\enddata
	\tablenotetext{a}{Maximum log-likelihood.}
	\tablenotetext{b}{$\du{\equiv}d/(4\,{\rm kpc})$}  
	\label{tab:fitsed_hess}
\end{deluxetable}

\end{document}